\documentclass[twocolumn,showpacs,preprintnumbers,nofootinbib,prd,superscriptaddress,groupedaddress,10pt]{revtex4-1}

\usepackage{graphicx,amssymb,amsmath,amsthm,amsfonts,epsfig,epsf}
\usepackage[linktocpage]{hyperref}
\usepackage[usenames]{color}
\usepackage{epstopdf}

\usepackage{aas_macros}
\usepackage{bm}
\usepackage{dcolumn}
\usepackage[latin1]{inputenc}
\usepackage{latexsym}
\usepackage{rotating}
\usepackage{hyperref}
\usepackage{color}
\usepackage{longtable}
\usepackage{enumerate}
\usepackage{tensor}
\usepackage{url}
\def\nn{\nonumber}

\def\be{\begin{equation}}
\def\ee{\end{equation}}
\newcommand{\beq}{\begin{eqnarray}}
\newcommand{\eeq}{\end{eqnarray}} 
\newcommand{\ba}{\begin{align}}
\newcommand{\ea}{\end{align}}

\begin{document}

\title{Asymptotically anti-de Sitter Proca Stars}

\author{
Miguel Duarte$^{1}$,
Richard Brito$^{1}$,
}
\affiliation{${^1}$ CENTRA, Departamento de F\'{\i}sica, Instituto Superior T\'ecnico -- IST, Universidade de Lisboa -- UL,
Avenida Rovisco Pais 1, 1049 Lisboa, Portugal}

\begin{abstract}
We show that complex, massive spin-1 fields minimally coupled to Einstein's gravity with a negative cosmological constant, admit asymptotically anti-de Sitter self-gravitating solutions. Focusing on 4-dimensional spacetimes, we start by obtaining analytical solutions in the test-field limit, where the Proca field equations can be solved in a fixed anti-de Sitter background, and then find fully non-linear solutions numerically. These solutions are a natural extension of the recently found asymptotically flat Proca stars and share similar properties with scalar boson stars. In particular, we show that they are stable against spherically symmetric linear perturbations for a range of fundamental frequencies limited by their point of maximum mass. We finish with an overview of the behavior of Proca stars in $5$ dimensions.
\end{abstract}



\maketitle

\section{Introduction}

Driven by the widely celebrated Anti-de Sitter/conformal field theory (AdS/CFT) correspondence conjecture~\cite{Maldacena:1997re,Witten:1998qj}, the study of asymptotically anti-de Sitter (aAdS) spacetimes has attracted a great deal of interest in the last two decades. One of the main defining aspects of an aAdS spacetime is the existence of a timelike boundary at conformal infinity, which requires suitable boundary conditions to be provided at infinity, if one wishes to have a well-posed initial value problem in a dynamical scenario~\cite{Friedrich:1995vb}. This property is the reason why AdS is normally referred to behave as a ``confining box'', and is the main reason for the existence of some very peculiar phenomena occurring in AdS. For example, dynamical studies of the propagation of scalar fields in AdS have led to the conclusion that AdS is nonlinearly unstable against black-hole formation for a wide class of smooth initial data, regardless of the field's initial amplitude~\cite{Bizon:2011gg,Jalmuzna:2011qw,Buchel:2012uh}. This is thought to be caused by the confinement property of the AdS boundary, which allows non-linear effects to build up over time.

However, it is now clear that there are also large classes of initial data which are able to avoid collapse to a BH by forming stable configurations. For scalar fields and pure  gravitational perturbations, these ``islands of stability'' seem to be closely related to self-gravitating time-periodic solutions of Einstein's field equations with a negative cosmological constant~\cite{Dias:2011ss,Dias:2012tq,Maliborski:2013jca,Horowitz:2014hja,Fodor:2015eia}, known as geons or oscillons, depending on whether they are made of gravitational radiation or real scalar fields, respectively. For complex scalar fields, analogous solutions known as boson stars can also be constructed, but the existence of a global $U(1)$ symmetry allows for the existence of solutions with a time-periodic scalar field, while keeping the metric static or stationary~\cite{Kaup:1968zz,Ruffini:1969qy,Astefanesei:2003qy,Buchel:2013uba}.

Boson stars are compact self-gravitating solutions of the Einstein-Klein-Gordon field equations which were first studied more than forty years ago for a vanishing cosmological constant~\cite{Kaup:1968zz,Ruffini:1969qy} (for a recent review see Ref.~\cite{Liebling:2012fv}).
On the other hand, boson stars in spacetimes with a negative cosmological constant were first constructed in Ref.~\cite{Astefanesei:2003qy} and recently re-explored in the context of the weakly turbulent instability of AdS~\cite{Buchel:2013uba}. These solutions were shown to be stable against sufficiently small perturbations and apparently immune to the weakly turbulent instability~\cite{Buchel:2013uba}.

Analogous configurations for spin-1 fields have seen far less progress than their spinless counterpart. Nonetheless, complex massive spin-1 fields minimally coupled to Einstein's gravity were recently shown to also allow for the existence of asymptotically flat self-gravitating solutions~\cite{Brito:2015pxa}. These solutions, dubbed Proca stars, were shown to share very similar properties to boson stars. In particular, a subset of solutions was shown to be linearly stable~\cite{Brito:2015pxa} and they are believed to be formed under generic initial conditions~\cite{Garfinkle:2003jf,Brito:2015yfh}. Moreover, the rotating Proca stars built in Ref.~\cite{Brito:2015pxa} were shown to be continuously connected to asymptotically flat rotating black holes with Proca hair~\cite{Herdeiro:2016tmi}, similarly to rotating boson stars~\cite{Herdeiro:2014goa}. Following these works, charged Proca stars were constructed in Ref.~\cite{Garcia:2016ldc}, while in the context of the AdS/CFT duality and holographic superfluids, Ref.~\cite{Arias:2016nww} (see also~\cite{Cai:2013aca}) studied hairy planar BH solutions of the Einstein-Maxwell-Proca field equations.

Here we show that by adding a negative cosmological constant, one can also construct Proca stars with AdS asymptotics, and that a subset of these solutions is linearly stable against radial perturbations. Similarly to the scalar field and gravitational cases~\cite{Dias:2011ss,Dias:2012tq,Buchel:2013uba}, in the test-field limit these solutions are related to the normal modes of a Proca field in AdS. Finally, we extend these results to $4+1$ dimensions and argue that a negative cosmological constant is necessary for the existence of stable Proca stars in higher-dimensional spacetimes.
		
\section{Setup}
\subsection{Framework}
We are interested in a complex Proca field\footnote{Note that a complex Proca field can also be described by two independent real vector fields with the same mass $\mu$~\cite{Herdeiro:2016tmi}.} with mass $\mu$, described by the potential 1-form $\mathcal{A}$ and the field strength $\mathcal{F}=d\mathcal{A}$. The minimal Einstein-Proca model is then given by:
	\begin{equation}\label{action_proca}
	S=\int d^{4}\sqrt{-g}\left(\frac{R-2\Lambda}{16\pi G}-\frac{1}{4}\mathcal{F}_{\alpha \beta}\bar{\mathcal{F}}^{\alpha \beta}-\frac{1}{2}\mu^2\mathcal{A}_{\alpha}\bar{\mathcal{A}}^{\alpha}\right)\,,
	\end{equation}
where $\Lambda$ is the cosmological constant and $\bar{\mathcal{A}}$ and $\bar{\mathcal{F}}$ are the complex conjugates of the potential and the field strength, respectively.	The Einstein-Proca field equations implied by this action are:
	\beq
	G_{\alpha \beta}&=&8\pi GT_{\alpha\beta}\,,\label{eq_einstein}\\
	\nabla_\alpha \mathcal{F}^{\alpha \beta}&=&\mu^2\mathcal{A}^\beta\,,\label{eq_proca}
	\eeq
	where the stress-energy tensor is given by
	\beq\label{Tmunu}
	T_{\alpha \beta}&=&-\mathcal{F}_{\sigma (\alpha}\bar{\mathcal{F}}_{\beta)}^{\sigma}-\frac{1}{4}g_{\alpha \beta}\mathcal{F}_{\sigma \tau}\bar{\mathcal{F}}^{\sigma \tau}\nn\\
	&+&\mu^2\left[\mathcal{A}_{(\alpha}\bar{\mathcal{A}}_{\beta)}-\frac{1}{2}g_{\alpha \beta}\mathcal{A}_{\sigma}\bar{\mathcal{A}}^{\sigma}\right]\,.
	\eeq
	One can easily check that, for $\mu\neq 0$, the Lorenz condition $\nabla_\alpha \mathcal{A}^\alpha=0$ is implied by the Proca field equation~\eqref{eq_proca}.

The action~\eqref{action_proca} is invariant under a global $U(1)$ symmetry of the form $\mathcal{A}_\mu\rightarrow e^{i\alpha}\mathcal{A}_\mu$, with $\alpha$ constant, which implies the existence of a conserved 4-current:
\be
j^\alpha=\frac{i}{2}\left(\bar{\mathcal{F}}^{\alpha \beta}\mathcal{A}_\beta-\mathcal{F}^{\alpha\beta}\bar{\mathcal{A}}_\beta\right)\,.
\ee
This in turn implies the existence of a conserved Noether charge $Q$ given by:
\begin{equation}
Q=-\int_\Sigma d^3x \sqrt{-g} j^t\,,
\label{q}
\end{equation}
where $j^t$ is the temporal component of the 4-current, and $\Sigma$ is a space-like hypersurface.

\subsection{Spherically symmetric ansatz and boundary conditions}

We consider spherically symmetric solutions with a metric of the form
\begin{equation}\label{an_1}
ds^2=-\sigma^2(r)F(r)dt^2+\frac{1}{F(r)}dr^2+r^2(d\theta^2+\sin^2{\theta}d\phi^2)\,,
\end{equation}
where $F(r)=1-2m(r)/r-\Lambda r^2/3$, and a time-periodic Proca potential given by
\begin{equation}\label{an_2}
\mathcal{A}=e^{-i\omega t}\left[f(r)dt+ig(r)dr\right]\,.
\end{equation}
Here, the functions $m(r)$, $f(r)$, $g(r)$ and $\sigma(r)$ are real functions of the radial coordinate only and $\omega$ is a real parameter. 

Denoting radial derivatives with a prime, Einstein's field equations~\eqref{eq_einstein} yield
\beq\label{fieldeq1}
m'&=&4\pi Gr^2\left[\frac{(f'-\omega g)^2}{2\sigma^2}+\frac{1}{2}\mu^2\left(g^2F+\frac{f^2}{F\sigma^2}\right)\right]\,,\\
\sigma'&=&4\pi Gr\mu^2\sigma\left(g^2+\frac{f^2}{F^2\sigma^2}\right)\,,\label{fieldeq2}
\eeq
while from the Proca field equations~\eqref{eq_proca} we obtain
\beq\label{fieldeq3}
\left[\frac{r^2(f'-\omega g)}{\sigma}\right]'&=&\frac{\mu^2r^2f}{\sigma F}\,,\\
\label{fieldeq4}
\omega g-f'&=&\frac{\mu^2\sigma^2Fg}{\omega}\,.
\eeq
It is worth noting that the equations are very similar to the ones obtained in Ref.~\cite{Brito:2015pxa}, where asymptotically flat Proca stars were studied, except for the addition of a cosmological constant term in the metric function $F(r)$.

For these solutions, the Noether charge~\eqref{q} can be computed using 
\begin{equation}\label{q_spheric}
Q=\frac{4\pi \mu^2}{\omega}\int^{\infty}_0dr\, r^2g(r)^2\sigma(r)F(r)\,,
\end{equation}
while their energy is given by
\begin{equation}\label{density_eq}
\rho=-T^t_t=
 \frac{(f'-\omega g)^2}{2\sigma^2}
+\frac{1}{2}\mu^2 \left(g^2F+\frac{f^2}{F\sigma^2} \right)\,.
\end{equation}
To solve this system of equations we impose regular boundary conditions both at the origin and at infinity. Close to $r=0$, the field equations~\eqref{fieldeq1}~--~\eqref{fieldeq4} imply that:
\begin{equation}\label{b1}
\begin{aligned}
&f(r)=f_{0}+\frac{f_{0}}{6}r^2 \left(\mu^2-\frac{\omega^2}{\sigma_{0}^2}\right)+\mathcal{O}(r^4),\\
&g(r)=-\frac{f_0\omega}{3\sigma_0^2}r+\mathcal{O}(r^3),\\
&m(r)=\frac{4\pi Gf_{0}^{2}\mu^2}{6\sigma_{0}^{2}}r^3+\mathcal{O}(r^5),\\
&\sigma(r)=\sigma_{0}+\frac{4\pi G f_{0}^2\mu^2}{2\sigma_{0}}r^2+\mathcal{O}(r^4),
\end{aligned}
\end{equation}
where $f_0$ and $\sigma_0$ are constants. On the other hand, when $r\to \infty$, the only regular solutions behave as
\begin{equation}\label{b2}
\begin{aligned}
&f(r)=c_0r^{\alpha}...,\\
&g(r)=-\frac{c_0l^4\omega}{\alpha+1}r^{\alpha-3}+...,\\
&m(r)=M+4\pi Gc_0^2\frac{\alpha^2+\mu^2l^2}{2(1+2\alpha)}r^{2\alpha+1}+...,\\
&\log\sigma(r)=4\pi Gc_0^2\frac{\mu^2l^4}{2(\alpha-1)}r^{2\alpha-2}+...,
\end{aligned}
\end{equation}
where $c_0$ is a constant, $M$ is the ADM mass, $l^2=-3/\Lambda$ is the AdS curvature radius squared and $\alpha=-(\sqrt{1+4\mu^2l^2}+1)/2$. Here we should note that, as expected, the cosmological constant changes the behavior of the Proca field at infinity, when compared to the asymptotically flat case where the Proca field decays exponentially~\cite{Herdeiro:2016tmi}.
%

\section{Analytical Solutions}

Before computing the fully non-linear Proca stars, let us gain some insight about the nature of these solutions by considering a perturbative expansion in the amplitude of the vector potential $\mathcal{A}$. These small-amplitude solutions are simply the normal modes of a Proca field in AdS, which for test scalar fields and pure gravitational perturbations have been shown to be related to boson stars and geons~\cite{Dias:2011ss,Buchel:2012uh,Buchel:2013uba}.

We start by considering an expansion of the form
\beq
f(r)&=&\epsilon \frac{u_1(r)}{r}+\mathcal{O}(\epsilon^3)\,,\nn\\
g(r)&=&\epsilon \frac{u_2(r)}{r F_0(r)}+\mathcal{O}(\epsilon^3)\,,\nn\\
m(r)&=& \epsilon^2 m_2(r)+\mathcal{O}(\epsilon^4)\,,\nn\\
\sigma(r)&=&1+\epsilon^2 \sigma_2(r)+\mathcal{O}(\epsilon^4)\,,\nn
\eeq
where $F_0(r)=1+\frac{r^2}{l^2}$, $l^2=-3/\Lambda$ and $\epsilon$ is a small-bookkeeping parameter, which can be chosen to be the free-parameter $f_0\equiv f(r=0)=\epsilon$, by normalizing the function $u_1$ as
\begin{equation}
\frac{u_1}{r}\biggr\rvert_{r\rightarrow 0}=1\,.
\end{equation}

At the linear order in $\epsilon$, the problem reduces to solving the Proca field equations in an AdS background. The Lorenz condition, $\nabla_\alpha A^\alpha=0$, implies that 
\be
u_1=-\frac{F_0\partial_r(r\,u_2)}{\omega r}\,.
\ee
From this condition and the $(r)$ component of the Proca field equations~\eqref{eq_proca}, we then obtain the following master equation:
\begin{equation}\label{procaeqads}
\frac{d^2u_2}{dr_*^2}+\left[\omega^2-F_0\left(\frac{2}{r^2}+\mu^2\right)\right]u_2=0\,,
\end{equation}
where we defined the tortoise coordinate $r_*$ as $d r_*=F_0^{-1}d r$.

To solve this equation we change the radial variable to $x=\sin^2(r_*/l)$ and define a new radial function $Z(x)$ as
\begin{equation}
u_2(x)=(1-x)^{(3-k)/2}x\,Z(x)\,.
\end{equation}
where $k=5/2+\sqrt{1/4+\mu^2l^2}$, such that eq.~\eqref{procaeqads} becomes a hypergeometric differential equation for $Z(x)$:
\begin{equation}
x(1-x)\frac{d^2Z}{d x^2}+\left[c-(a+b+1)x\right]\frac{d Z}{d x}-(ab)Z=0\,,
\end{equation}
where $a=(5-k-l\omega)/2$, $b=(5-k+l\omega)/2$ and $c=5/2$. The most general solution to this equation is given by
\beq
&&Z(x)=A\, {_2}F_1\left(a,b;c;x\right)+\\\nn
&&B x^{1-c}\, {_2}F_1\left(1+a-c,1+b-c;2-c;x\right)\,.
\eeq
Requiring a regular solution at the origin $r=0$, implies that $B=0$, while imposing regularity of the solution at infinity gives the spectrum:
\begin{equation}\label{procaspectrum}
\omega l=2n+k\,,
\end{equation}
where $n=\{0,1,2,\dots\}$.
	
The backreaction of these solutions on the metric can be obtained by considering higher-order corrections in $\epsilon$. In particular, at order $\epsilon^2$, the ADM mass $M$ of these solutions can be obtained by integrating the radial equation for $m_2(r)$ obtained from expanding eq.~\eqref{fieldeq1}. For the fundamental solution, $n=0$, we obtain:
\begin{equation}\label{massalambda2}
M/l= 4\pi\epsilon^2 \frac{\sqrt{\pi } k(k-3) (k-2) \Gamma\left[k-\frac{3}{2}\right]}{24 \Gamma[k]}\,,
\end{equation}
while the charge can be obtained from~\eqref{q_spheric}, and reads:
\begin{equation}\label{charge_ana}
Q/l^2= 4\pi\epsilon^2 \frac{\sqrt{\pi } (k-3) (k-2) \Gamma\left[k-\frac{3}{2}\right]}{24 \Gamma[k]}\,.
\end{equation}
In this small-amplitude limit, we also find that for any $n$ the charge and the mass are related by
\be\label{MvsQ}
M=\omega \,Q\,,
\ee
with $\omega$ given by eq.~\eqref{procaspectrum}.
 
In the next Section, we will show that these results agree very well with fully non-linear solutions of the system of equations~\eqref{fieldeq1}~--~\eqref{fieldeq4}.

\section{Numerical Solutions}\label{sec:num}
\begin{figure*}[htb!]
\begin{center}
\begin{tabular}{cc}
\epsfig{file=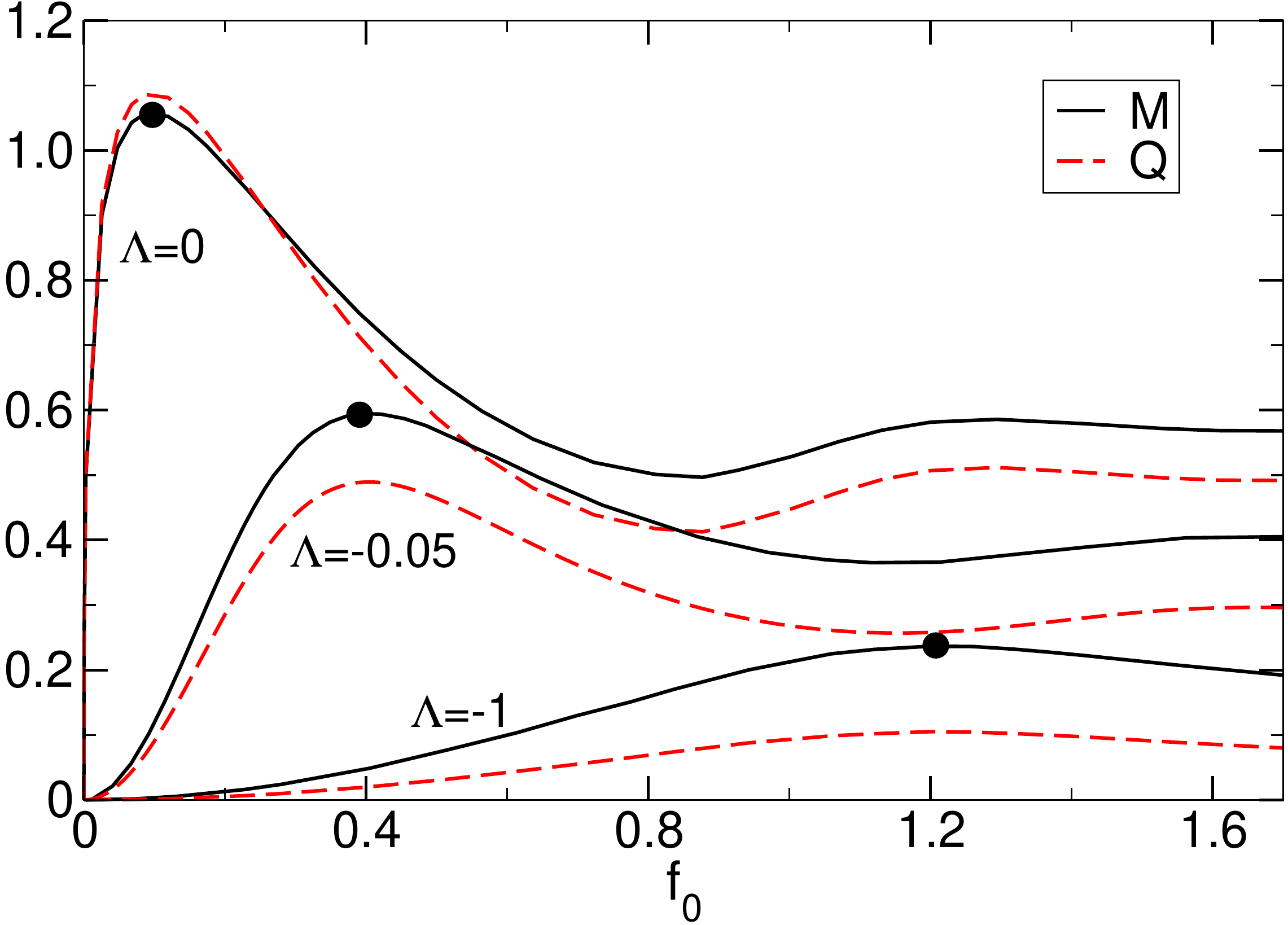,width=7.5cm,angle=0,clip=true}&
\epsfig{file=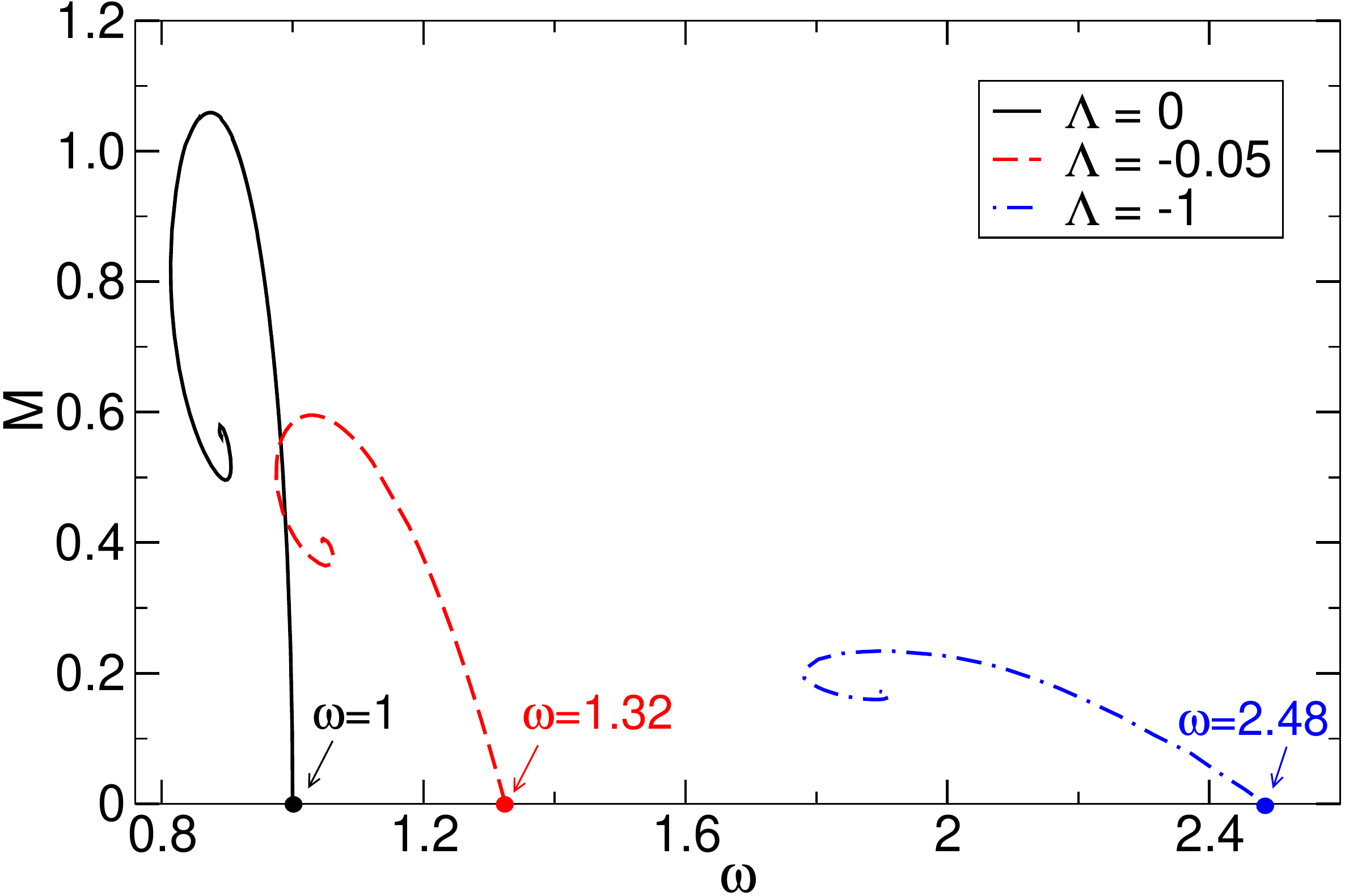,width=7.5cm,angle=0,clip=true}
\end{tabular}
\caption{Left panel: ADM mass $M$ and charge $Q$ as a function of the parameter $f_0$ for three different values of $\Lambda$. The black dots denote the point of maximum mass $M_{\rm max}$. Right panel: ADM mass $M$ as a function of the fundamental frequency $\omega$ for three different values of $\Lambda$.\label{massalambdas}}
\end{center}
\end{figure*}

\begin{figure}[ht!]
\begin{center}
\begin{tabular}{c}
\epsfig{file=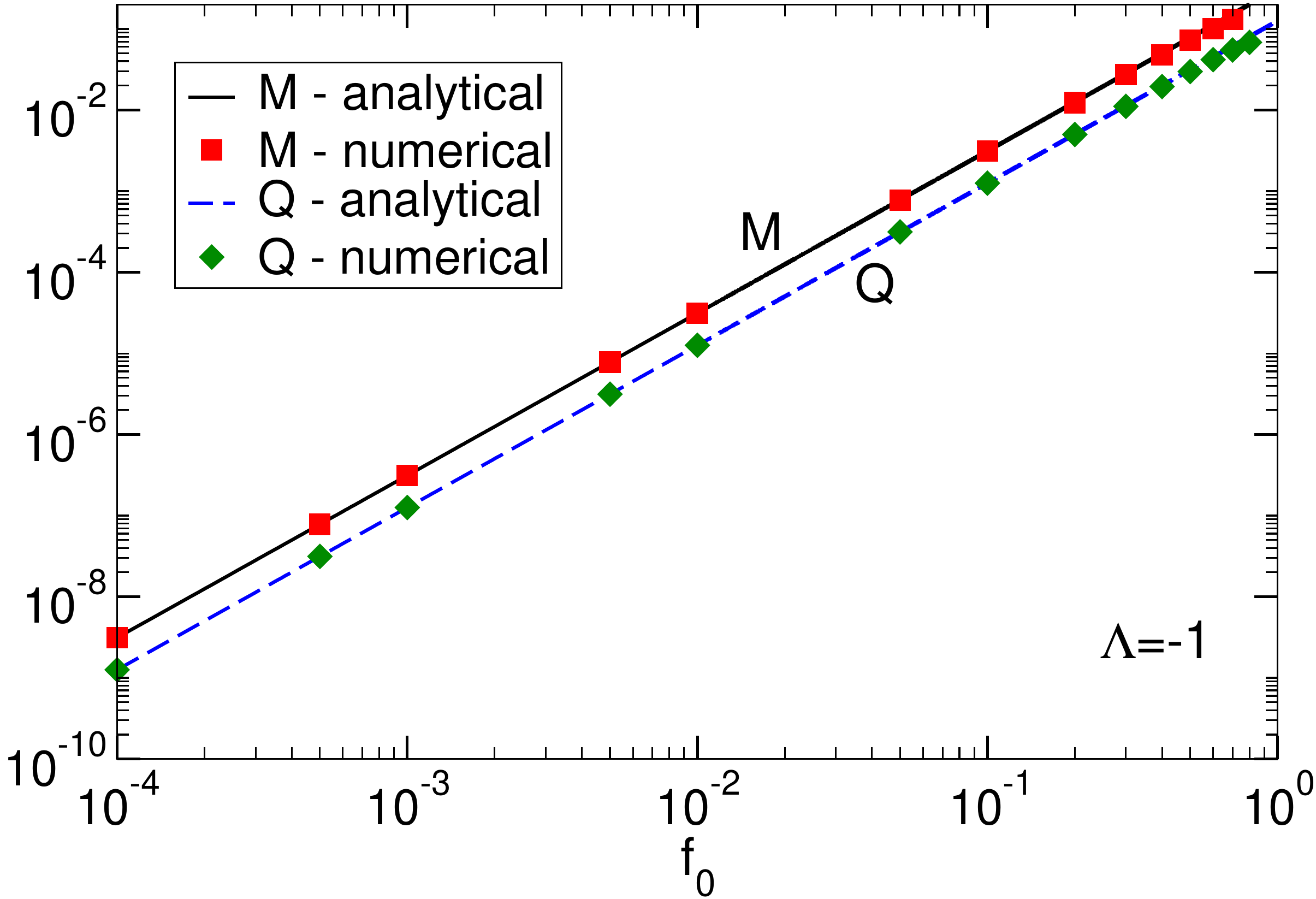,width=7.5cm,angle=0,clip=true}
\end{tabular}
\caption{Comparison between the perturbative approximations for the mass~\eqref{massalambda2} and charge~\eqref{charge_ana} and the fully non-linear results, as a function of $f_0$ for $\Lambda=-1$.\label{compareproca}}
\end{center}
\end{figure}

\begin{figure}[htb!]
\begin{center}
\begin{tabular}{c}
\epsfig{file=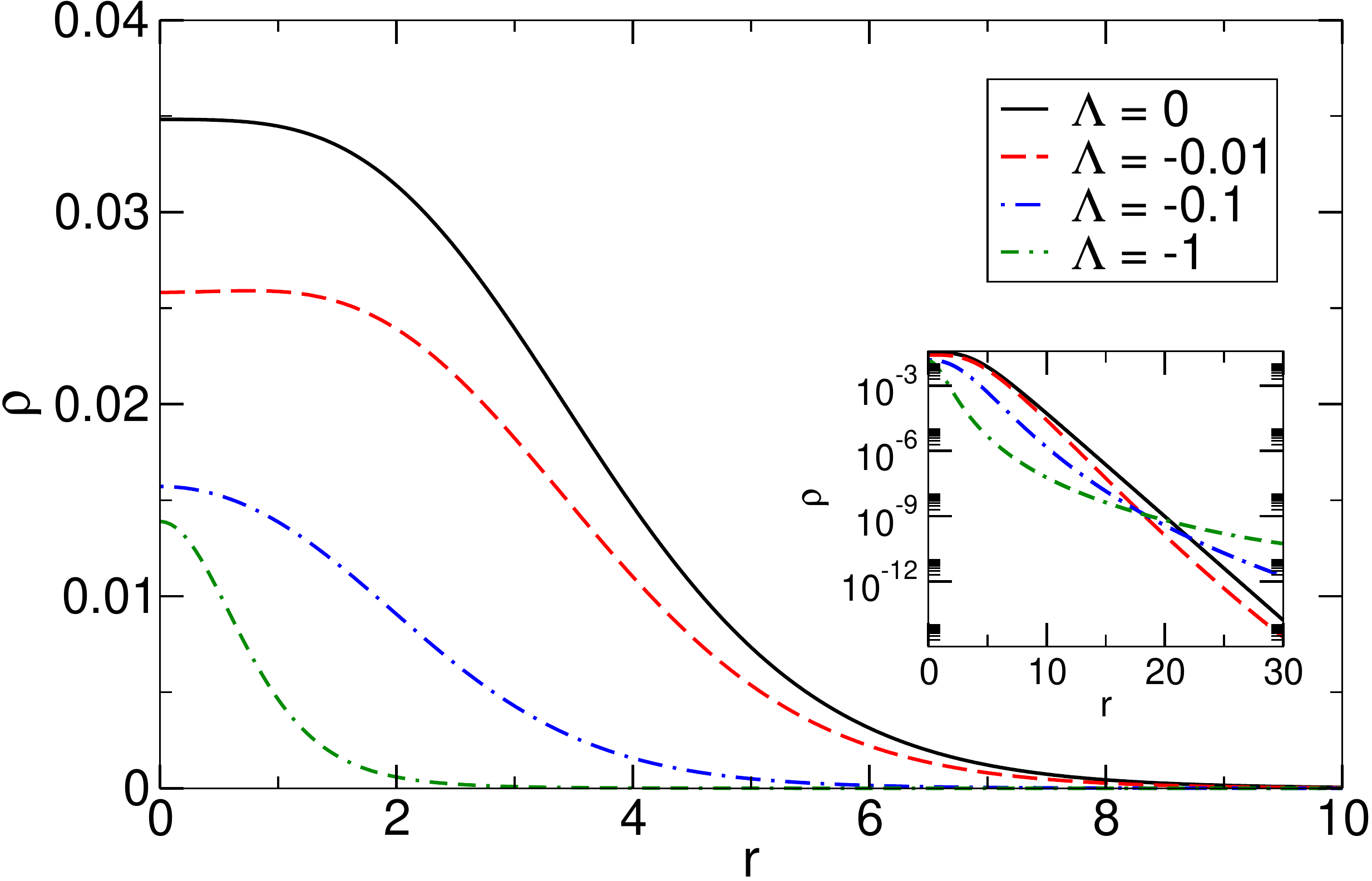,width=7.5cm,angle=0,clip=true}
\end{tabular}
\caption{Energy density $\rho=-T_t^t$ of the star for $f_0=0.165$ and different values of $\Lambda$. The inset plot shows a zoom of the solutions for large $r$.\label{density}}
\end{center}
\end{figure}

We are now in position to find the fully non-linear aAdS Proca stars. The set of ODEs~\eqref{fieldeq1}~--~\eqref{fieldeq4} can be solved numerically to find solutions fulfilling the boundary conditions~\eqref{b1} and~\eqref{b2}. 
For a given choice of the parameter $f_0$, we use the initial conditions~\eqref{b1} and numerically integrate the equations from $r=10^{-3}$ towards large $r$, and adjust the shooting parameter $\omega$, to find the solutions satisfying the boundary condition~\eqref{b2}. Fixing $\Lambda$ and varying $f_0$ we then find a set of solutions which are summarized in Figs.~\ref{massalambdas}~--~\ref{density}. We should note that for a given $\Lambda$ and $f_0$ there is an infinite family of solutions parametrized by the number of nodes in the profile of the function $g(r)$. For concreteness, here we focus on the fundamental solutions, for which $g(r)$ is nodeless. For numerical purposes, and in the following, we set $\mu=G=1$ by using the rescalings $r\to r \mu$, $m\to m \mu$, $\omega\to \omega/ \mu$, $\Lambda\to \Lambda/\mu^2$ and $f\to f \sqrt{4\pi G}$, $g\to g \sqrt{4\pi G}$.

In the left panel of Fig.~\ref{massalambdas} we show the ADM mass $M$ and the charge $Q$ of the fundamental solutions as a function of $f_0$ for three different values of the cosmological constant, while in the right panel we show the ADM mass as a function of the fundamental frequency $\omega$ for the same values of $\Lambda$.

For $\Lambda=0$ our results agree with the ones obtained in Ref.~\cite{Brito:2015pxa}. In particular, we obtain a maximum ADM mass given by $M_{\rm max}\simeq 1.058$, while the fundamental frequency is limited to a range of frequencies satisfying $\omega\lesssim 1$. In the limit $\omega\to 1$ the solutions become spatially diluted, while their mass and Noether charge vanish, but $M/Q\to 1$. 

On the other hand for $\Lambda\neq 0$, the eigenfrequencies of the solutions approach the spectrum given by eq.~\eqref{procaspectrum} when $f_0\to 0$, while their typical size\footnote{Proca stars do not have a hard surface and a well-defined radius. A typical definition for the effective radius of the star is to consider the radius containing, e.g., $\sim 99\%$ of the star's mass.} is always of the order $R\sim 10/\omega$.
For small $f_0$, their ADM mass and charge are well-described by~\eqref{massalambda2} and~\eqref{charge_ana}, as shown in Fig.~\ref{compareproca}, where we compare our numerical results for $\Lambda=-1$ with the perturbative approximations of the previous Section\footnote{Note that due to the rescaling employed in this Section we have $\epsilon=f_0/(\sqrt{4\pi})$.}. 

Overall, the behavior of the ADM mass when increasing $f_0$ is qualitatively similar to the asymptotically flat case. There is always a maximum mass, not foreseen by the linear analysis of the last Section, which is a monotonically decreasing function of $|\Lambda|$. For example, for $\Lambda=-0.05$ we obtain $M_{\rm max}\simeq 0.595$, while for $\Lambda=-1$ we find $M_{\rm max}\simeq 0.234$. The location of the maximum mass $M_{\rm max}$ for each $\Lambda$ is denoted by a black dot in the left panel of Fig.~\ref{massalambdas}. As we will show in the next Section, solutions to the left of this point are stable against radial perturbations while every solution to the right is unstable. 

%

In Fig.~\ref{density} we compare the typical density profile $\rho=-T_t^t$, given by eq.~\eqref{density_eq}, for solutions with different values of the cosmological constant $\Lambda$. As expected, for small $|\Lambda|$, aAdS Proca stars are nearly indistinguishable from the asymptotically flat ones. In the inset plot, we can see that at intermediate distances $r$, the small-$|\Lambda|$ solutions are dominated by the typical exponential decay $\rho\sim e^{-2r\sqrt{\mu^2-\omega^2}}$ (cf. dashed red curve in Fig.~\ref{density}), typically found in asymptotically flat Proca stars~\cite{Brito:2015pxa} (cf. solid black curve in Fig.~\ref{density}). For these solutions, the cosmological constant only starts to dominate at large $r$, with the field behaving as~\eqref{b2}. On the other hand, for large $|\Lambda|$, the effect of the mass term is always sub-dominant compared to the cosmological constant term and the solutions differ considerably from the asymptotically flat ones.

Finally, we note that when $\mu=0$, spherically symmetric regular solutions do not exist, for any value of $\Lambda$. This is consistent with the fact that Maxwell fields do not allow for spherically symmetric solitonic configurations, but is in contrast with the scalar field case, for which boson stars exist for $\mu=0$, but only when $\Lambda<0$~\cite{Buchel:2013uba}.
This behavior can be understood from the fact that the solutions discussed here are analogous to stationary waves in a box. Since Maxwell fields do not allow for everywhere regular spherically symmetric waves, then solitonic solutions cannot exist when $\mu=0$. 

\section{Stability against linear perturbations}\label{stab}

Having established the existence of Proca stars with AdS asymptotics, we now study the stability of these solutions against small radial perturbations. It has been shown, for both scalar boson stars~\cite{Gleiser:1988ih,Lee:1988av} and Proca stars~\cite{Brito:2015pxa}, that the asymptotically flat solutions are stable from the point $f_0\to 0$, where $M\to 0$, up to the value $f_0^{\rm c}$ corresponding the maximal mass $M_{\rm max}$ (cf. black dots in Fig.~\ref{massalambdas}). Similar studies have shown that the same rule applies for aAdS boson stars~\cite{Astefanesei:2003qy,Buchel:2013uba}, and this picture is also supported by fully non-linear evolutions of the field equations for aAdS boson stars~\cite{Buchel:2013uba}. Here we will show that the same conclusion can be drawn for aAdS Proca stars.

We consider linear perturbations around the ground state of the spherically symmetric Proca stars, assuming that all perturbations have a harmonic time dependence of the form $e^{-i\Omega t}$, with $\Omega$ the characteristic vibrational frequencies. Following~\cite{Brito:2015pxa}, the perturbed metric can be written as
\beq\label{pertmetric}
ds^2&=&-\sigma^2(r)F(r)[1-\epsilon H_0(r)e^{-i\Omega t}]dt^2\nn\\
&+&\frac{[1+\epsilon H_2(r)e^{-i\Omega t}]}{F(r)}dr^2+r^2d\Omega ^2\,,
\eeq	
while the perturbed vector field can be written as
\beq\label{pertfield}
&&\mathcal{A}=e^{-i\omega t}\left[\left(f(r)+e^{-i\Omega t}\frac{\epsilon f_1(r)+i \epsilon f_2(r)}{r}\right)dt\right.\nn\\
&&\left.+\left(ig(r)+e^{-i\Omega t}\frac{\epsilon g_1(r)+i \epsilon g_2(r)}{r}\right)dr\right]\,,\nn\\
&&\bar{\mathcal{A}}=e^{i\omega t}\left[\left(f(r)+e^{-i\Omega t}\frac{\epsilon f_1(r)-i \epsilon f_2(r)}{r}\right)dt\right.\nn\\
&&\left.+\left(-ig(r)+e^{-i\Omega t}\frac{\epsilon g_1(r)-i \epsilon g_2(r)}{r}\right)dr\right]\,,
\eeq
where $\epsilon$ is a small, bookkeeping parameter and $H_0$, $H_2$, $f_1$, $f_2$, $g_1$ and $g_2$ are radial perturbations around the background solutions.

As shown in Appendix~\ref{app:stability}, the perturbed Einstein-Proca field equations~\eqref{eq_einstein} and~\eqref{eq_proca} can be reduced to a set of coupled ODE's. To solve this system of equations we impose regularity of the perturbations at the origin and at infinity.
At $r=0$, we get the following boundary conditions for the perturbed quantities 
	\begin{equation}
	\begin{aligned}
	&H_0(r)=h_0+\mathcal{O}(r^2),\\
	&H_2(r)=\mathcal{O}(r^2),\\
	&f_1(r)=h_1r+\mathcal{O}(r^3),\\
	&f_2(r)=h_2r+\mathcal{O}(r^3),\\
	&g_2(r)=\mathcal{O}(r^2),
	\end{aligned}
	\end{equation}
where the coefficients $h_0$, $h_1$ and $h_2$ are constants. Using the linearity of the system of equations~\eqref{perteq1}--~\eqref{perteq5}, we can set $h_1=1$. We see from~\eqref{perteq4} that, when $\Omega=0$, the equation for $f_2$ decouples. Thus, we can set $h_2=0$ and check a posteriori that this is consistent with the boundary conditions. At infinity the system of equations~\eqref{perteq1}~--~\eqref{perteq5} allows for the following boundary conditions:
\be\label{boundary_infty}
X\to 0\,,\quad \rm{as}\quad r\to\infty\,,
\ee
where $X$ collectively denotes $H'_0$, $H_2$, $f_1$, $f_2$ and $g_2$. We checked that imposing two of these conditions is sufficient to ensure all the others. The system~\eqref{perteq1}--~\eqref{perteq5}, jointly with these boundary conditions is then a 2-dimensional eigenvalue problem for $\Omega$ and $h_0$, which can be solved using the same shooting method outlined in the previous Section and described in Refs.~\cite{Hawley:2002zn,Brito:2015pxa}. 

\begin{figure}[htb!]
\begin{center}
\begin{tabular}{c}
\epsfig{file=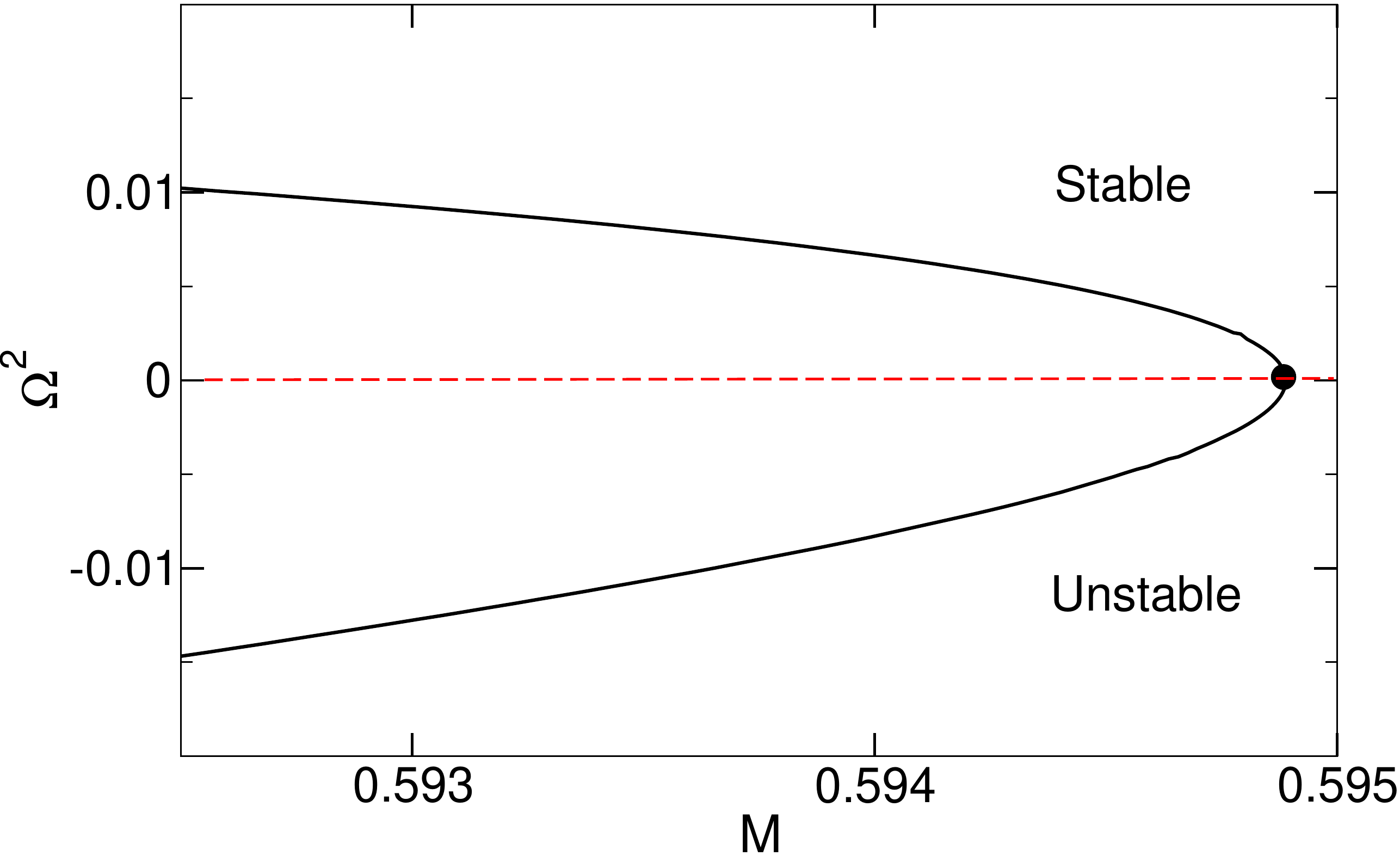,width=7.5cm,angle=0,clip=true}
\end{tabular}
\caption{The squared characteristic vibrational frequency as a function of the total mass of the Proca star for $\Lambda=-0.05$. The critical mass at which the star becomes unstable is also its point of maximum mass, denoted by a black dot.\label{stability}}
\end{center}
\end{figure}

The fundamental characteristic frequency $\Omega$ as a function of the Proca star's mass is shown in Fig.~\ref{stability} for $\Lambda=-0.05$, where the black dot denotes the point of maximum mass $M_{\rm max}\sim 0.595$ shown in the left panel of Fig.~\ref{massalambdas}. A similar behavior can be found for other values of $\Lambda$. 
We can see that $\Omega$ is a real number for $f_0<f_0^{\rm c}$, which corresponds to stable normal modes of the star. For $f_0>f_0^{\rm c}$, $\Omega$ becomes a pure positive imaginary number, and thus from~\eqref{pertmetric} and~\eqref{pertfield} we conclude that these solutions are unstable.

Unlike in the asymptotically flat case, where unstable solutions can migrate to the stable branch via mass ejection~\cite{Seidel:1990jh,Seidel:1993zk,Alcubierre:2003sx,Brito:2015yga,Brito:2015yfh} or simply completely disperse, the confinement property of the AdS boundary makes these scenarios unlikely to occur for aAdS solutions. Although fully non-linear evolutions are needed to access the end-state of the unstable solutions, the most likely scenario is that they quickly collapse to black holes.

Finally, we note that for asymptotically flat solutions, one can typically define a ``binding energy'' through the quantity $B=1-M/(\mu Q)$ and show that solutions with $B<0$ are unstable~\cite{Gleiser:1988,Brito:2015pxa}. As can be seen in Fig.~\ref{massalambdas} and from eq.~\eqref{MvsQ}, for a negative cosmological constant we generically have $B<0$. As shown in this Section and in Ref.~\cite{Astefanesei:2003qy,Buchel:2013uba} for aAdS scalar boson stars, in aAdS spacetimes this does not necessarily imply that these solutions are unstable. Interestingly, if one defines the quantity $\tilde{B}=1-M/(\omega_{0} Q)$, where $\omega_{0}$ is the eigenfrequency of the solution in the limit $f_0\to 0$, we find that for any $\Lambda$, $\tilde{B}$ is always positive for $f_0<f_0^{\rm c}$, while it becomes negative for some value larger than $f_0^{\rm c}$, similarly to what happens for $\Lambda=0$. It is thus tempting to speculate that this quantity generalizes the concept of binding energy for a negative cosmological constant, and that solutions with $\tilde{B}<0$ are necessarily unstable.

\section{Solutions in Five Dimensions}
Solutions for Proca stars with or without a cosmological constant can easily be extended to higher-dimensional spacetimes. For concreteness, let us focus on five dimensions, although we expect the same behavior to occur for any dimension larger than four. For this case the spherically symmetric metric is given by
\begin{equation}
ds^2=\sigma^2(r)F(r)dt^2+\frac{1}{F(r)}dr^2+r^2d\Omega_3^2\,,
\end{equation}
where $d\Omega_3^2$ denotes the $3$-dimensional unit sphere line element and the function $F(r)$ is now given by
\begin{equation}
F(r)=1-\frac{2m(r)}{r^2}-\frac{\Lambda r^2}{6}\,.
\end{equation}
For the Proca potentials we keep the ansatz~\eqref{an_2}. Similarly to the previous Sections, Einstein's equations, together with the Proca field equations, yield a system of four ordinary differential equations that generalize~\eqref{fieldeq1}--~\eqref{fieldeq4}, which can be solved using the same methods of Section~\ref{sec:num}.

\begin{figure}[ht!]
\begin{center}
\epsfig{file=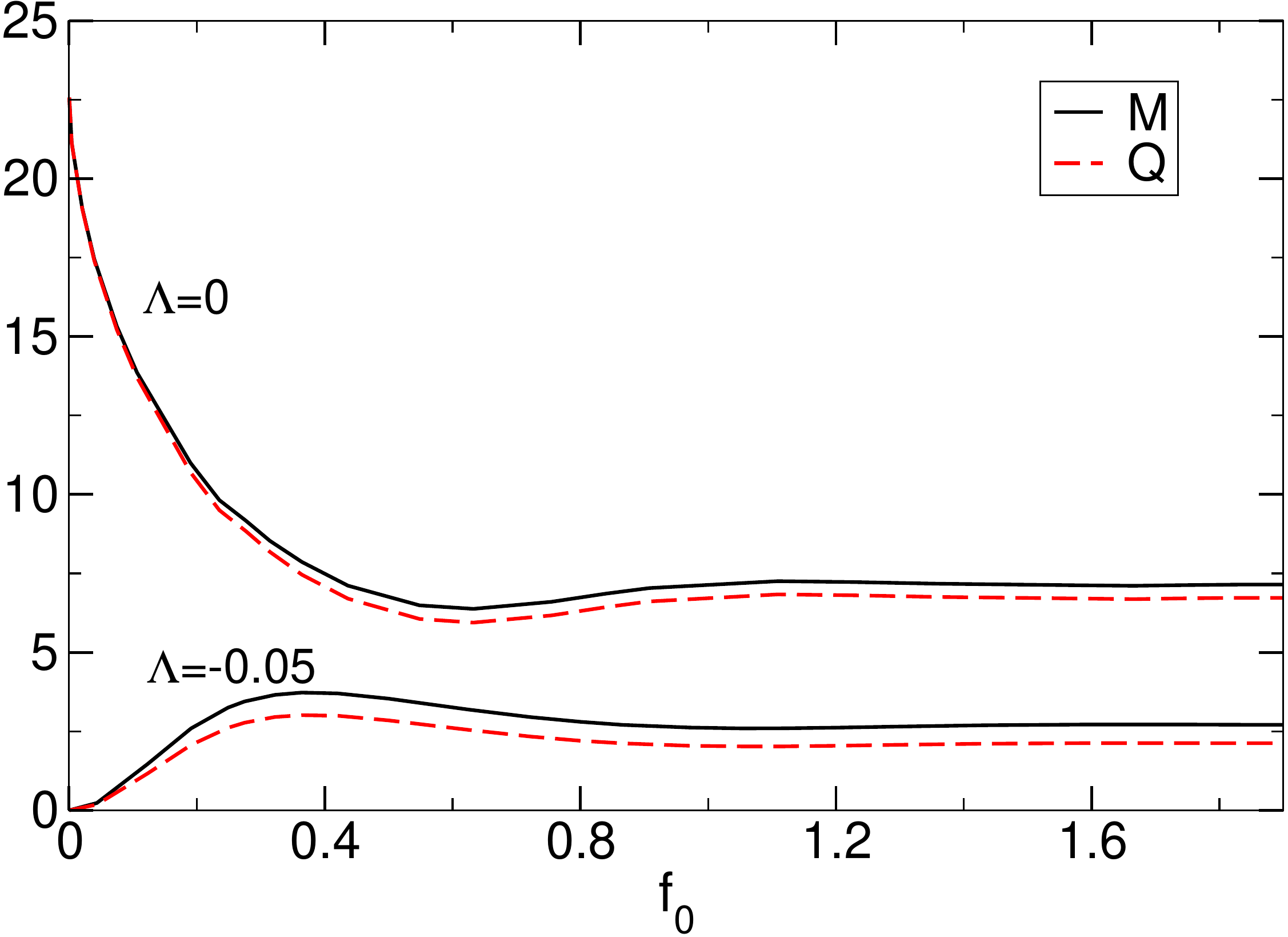,width=7.5cm,angle=0,clip=true}
\caption{ADM mass and charge of a 5-dimensional Proca star as a function of the parameter $f_0$, for $\Lambda=0$ and $\Lambda=-0.05$.\label{massdensityprocaflatads5d}}
\end{center}
\end{figure}

The main difference with the 4-dimensional case occurs when $\Lambda=0$. In 5 dimensions the mass and charge of the solutions go to a finite value $M=Q\sim 22.6$ in the limit $f_0\to 0$. This is shown in Fig.~\ref{massdensityprocaflatads5d}. This behavior can also be found for scalar boson stars~\cite{Hartmann:2010pm,Brihaye:2015jja} and seems to be unique to asymptotically flat solutions, as can be seen in Fig.~\ref{massdensityprocaflatads5d}.

We note that, unlike the 4-dimensional case discussed above, in 5 dimensions we find that for $\Lambda=0$, the binding energy is always negative, $1-M/(\mu Q)<0$, for any value of $f_0$. This behavior was also noticed for asymptotically flat scalar boson stars in Ref.~\cite{Brihaye:2015jja}, where it was argued that asymptotically flat 5D boson stars are always unstable. A dynamical analysis, similar to the one employed in the previous Section, confirms that asymptotically flat 5-dimensional boson stars are indeed always unstable against radial perturbations~\cite{Edgardo}.
We thus conclude that asymptotically flat 5-dimensional Proca stars should also be always unstable.
 
On the other hand, from Fig.~\ref{massdensityprocaflatads5d} it is also clear that for $\Lambda<0$, the mass and charge of the solutions vanish when $f_0\to 0$, while there is maximum mass and charge at a finite value of $f_0$, similarly to the 4-dimensional case (cf. Fig~\ref{massalambdas}). We thus expect that for $\Lambda<0$ there will always be a region of stable solutions. Unfortunately, due to the more complex nature of the 5-dimensional solutions, we were not able to confirm this instability using the same method of the previous Section. We thus leave a complete analysis of these solutions for future work.

\section{Conclusions}
We showed that complex massive spin-1 fields coupled to Einstein's gravity with a negative cosmological constant can form solitonic structures. These structures are qualitatively similar to aAdS scalar boson stars discussed in Refs.~\cite{Astefanesei:2003qy,Buchel:2013uba} and are a direct extension of the asymptotically flat Proca stars recently found in Ref.~\cite{Brito:2015pxa}.

Dynamical studies of matter fields in AdS have been mostly focused on scalar fields~\cite{Bizon:2011gg,Jalmuzna:2011qw,Buchel:2012uh,Maliborski:2013jca,Okawa:2015xma,Okawa:2014nea}. As shown here, it is very likely that the dynamics of massive vector fields in AdS will share very similar properties. For example, we expect that aAdS Proca stars will be another example of solutions able to avoid the weakly turbulent instability of confined geometries~\cite{Bizon:2011gg,Jalmuzna:2011qw,Buchel:2012uh,Dias:2011ss,Maliborski:2013jca}. 

There are various possible extensions of this work. A generalization of these solutions to real vector fields, analogous to the real scalar field oscillons~\cite{Maliborski:2013jca,Fodor:2015eia}\footnote{In asymptotically flat spacetimes these solutions are more frequently known as oscillatons~\cite{Seidel:1991zh,Brito:2015yfh}.}, should exist. We also expect that going beyond spherical symmetry one should also be able to construct rotating Proca stars, and it is very likely that the field equations allow for the existence of time-periodic \emph{massless} spin-1 geons, analogous to the gravitational geons built in Refs.~\cite{Dias:2011ss,Horowitz:2014hja}.

Furthermore, going beyond spherical symmetry might unveil a much richer structure. In fact, it was recently shown that Maxwell fields, i.e., massless spin-1 fields, allow for the existence of non-spherically symmetric static or stationary regular solutions with AdS asymptotics~\cite{Herdeiro:2015vaa,Costa:2015gol,Herdeiro:2016xnp}. As far as we know, whether a generalization of these solutions for massive spin-1 fields exists is currently unknown.

A further generalization of this work is the construction of aAdS Kerr black holes with Proca hair~\cite{Herdeiro:2016tmi}. Although one can prove that the only spherically symmetric black hole solution allowed by the field equations~\eqref{fieldeq1}--~\eqref{fieldeq4} is the Schwarzschild-AdS spacetime\footnote{A detailed proof is shown in~\cite{Herdeiro:2016tmi} for an asymptotically flat geometry, but the same procedure can easily be extended to AdS with the same result.}, we naturally expect that Kerr-AdS black holes with Proca hair, analogous to the solutions found in Ref.~\cite{Dias:2011at,Herdeiro:2014goa,Dias:2015rxy,Herdeiro:2016tmi}, should also exist. Additionally, one might also expect that other hairy black-hole solutions, analogue to the ones found in Ref.~\cite{Costa:2015gol,Herdeiro:2016plq} for Einstein-Maxwell theory with a negative cosmological constant, might also exist when adding a mass term to the vector field. We thus hope that this work will trigger more research in this direction.


\begin{acknowledgments}
We thank Vitor Cardoso for useful discussions and suggestions.
R.B. acknowledges financial support from the FCT-IDPASC program through the grant SFRH/BD/52047/2012. We acknowledge financial support provided under the European Union's H2020 ERC Consolidator Grant ``Matter and strong-field gravity: New frontiers in Einstein's theory'' grant agreement no. MaGRaTh-646597, and under the H2020-MSCA-RISE-2015 Grant No. StronGrHEP-690904.
\end{acknowledgments}

\appendix
\onecolumngrid
\section{Linear stability analysis: field equations\label{app:stability}}
Expanding the metric and the vector field as~\eqref{pertmetric} and~\eqref{pertfield}, and considering only linear terms in $\epsilon$, we find that the $(tr)$ component of the Einstein equations~\eqref{eq_einstein} gives
\begin{equation}\label{eqg1}
g_1=-\frac{2g\mu^2f_2+i\Omega H_2}{2\mu^2f}\,.
\end{equation}
Replacing this in the $(tt)$ component we get:
	\begin{equation}\label{perteq1}
	\begin{aligned}
	H_2'=&H_2\left(\frac{r\mu^2f^2}{F^2\sigma^2}-\frac{1}{rF}-\frac{\Omega^2g}{\omega f}+\frac{\Lambda r}{F}\right)+H_0\left(\frac{r\mu^2f}{F^2\sigma^2}+\frac{r\mu^4g^2F\sigma^2}{\omega^2}\right)+4g\mu^2g_2+f_2\frac{2i\mu^2\Omega g^2}{\omega f}+\\
	&2\mu^2f_1\left(\frac{g}{r\omega}+\frac{f}{\sigma^2F^2}\right)-f_1'\frac{2\mu^2g}{\omega}\,,
	\end{aligned}
	\end{equation}
	Multiplying the $(rr)$ component by $F^2\sigma^2$ and adding the $(tt)$ component one finds
	\begin{equation}
	H_0'=H_2'-4g\mu^2g_2-(H_0+H_2)\frac{2r\mu^2f^2}{F^2\sigma^2}-f_1\frac{4\mu^2f}{F^2\sigma^2}\,.
	\end{equation}
The $(r)$ component of the Proca field equations~\eqref{eq_proca} and its complex conjugate are independent and give
	\begin{equation}
	\begin{aligned}
	f_1'=&\frac{f_1}{r}+g_2\omega \left(1-\frac{\mu^2F\sigma^2}{\omega^2-\Omega^2}\right)+H_0\frac{\mu^2rgF\sigma^2(2\omega^2-\Omega^2)}{2\omega(\omega^2-\Omega^2)}+f_2\frac{i\Omega g(\mu^2F\sigma^2+\omega^2-\Omega^2)}{f(\omega^2-\Omega^2)}+\\
	&\frac{H_2\Omega^2}{2f}\left[\frac{F\sigma^2(\mu^2rfg-\omega)}{\omega^3-\omega\Omega^2}-\frac{1}{\mu^2}\right]\,,
	\end{aligned}
	\end{equation}
	\begin{equation}\label{perteq4}
	\begin{aligned}
	f_2'=&-i\Omega g_2\omega \left(1+\frac{\mu^2N\sigma^2}{\omega^2-\Omega^2}\right)+H_0\frac{i\Omega \mu^2rgF\sigma^2}{2(\omega^2-\Omega^2)}+f_2\left[\frac{1}{r}+\frac{\omega g(\omega^2-\Omega^2-\mu^2F\sigma^2)}{f(\omega^2-\Omega^2)}\right]+\\
	&\frac{iH_2\Omega}{2f}\left[\frac{F\sigma^2(\mu^2rfg-\omega)}{\omega^2-\Omega^2}+\frac{\omega}{\mu^2}\right]\,.
	\end{aligned}
	\end{equation}
Finally, from the $(t)$ component of the Proca equations we find
	\begin{equation}\label{perteq5}
	\begin{aligned}
	g_2'=&-f_1\frac{\omega}{F^2\sigma^2}+g_2\left[\frac{\mu^4rg^2F\sigma^2(3\omega^2-\Omega^2)}{\omega^2(\omega^2-\Omega^2)}-\frac{1}{rF}+\frac{r\Lambda}{F}\right]+H_0\left(-\frac{r\omega f}{F^2\sigma^2}-\frac{\mu^4r^2g^3F\sigma^2}{\omega^2-\Omega^2}\right)+\\
	&H_2\left[\frac{\mu^2r\Omega^2g^2F\sigma^2(\omega-\mu^2rfg)}{\omega^2f(\omega^2-\Omega^2)}-\frac{r\omega f}{F^2\sigma^2}-\frac{g}{F}+\frac{r^2\Lambda g}{F}\right]+\frac{i\Omega f_2}{F^2\sigma^2}\left(\frac{2\mu^4rg^3F^3\sigma^4}{\omega^3f-\omega\Omega^2f}+1\right)\,.
	\end{aligned}
	\end{equation}
	%
\twocolumngrid
\bibliographystyle{h-physrev4}
\bibliography{ref}
	
\end{document}